\documentclass[11pt, oneside]{article}   	
\usepackage{geometry}                		
\usepackage{graphicx}				
\usepackage{amssymb,amsmath}
\usepackage{wrapfig}

\title{Investigation of \textsf{Skyscraper}'s Feat \\[-2mm] {\small (original version with an addendum)}}
\author{Costas Efthimiou\footnote{Contact email: costas@physics.ucf.edu}\\ \small Department of Physics\\ 
\small University of Central Florida \\ \small Orlando, FL 32816}

\date{}

\begin{document}
\maketitle

\begin{abstract}
\textsf{Skyscraper} is a Hollywood action  film directed and written by Rawson M. Thurber scheduled to be released on July 13, 2018. 
We present an exhaustive analysis of the feat shown in the recently released teaser poster and trailer  of the film. Although the feat
appears to be unrealistic at first glance, after close investigation using back-of-the-envelope calculations, it is seen to be within human 
capabilities.

This article is the original version of an abridged article published in \textsf{Physics Education} \cite{Efthimiou}. It was written very soon after the poster and clip were released by Universal Pictures. I have left intact the original material. (See addendum.)
\end{abstract}

\newpage
\section{Introduction}

\begin{wrapfigure}[23]{r}[0pt]{5cm}
\begin{center}
\includegraphics[height=8cm]{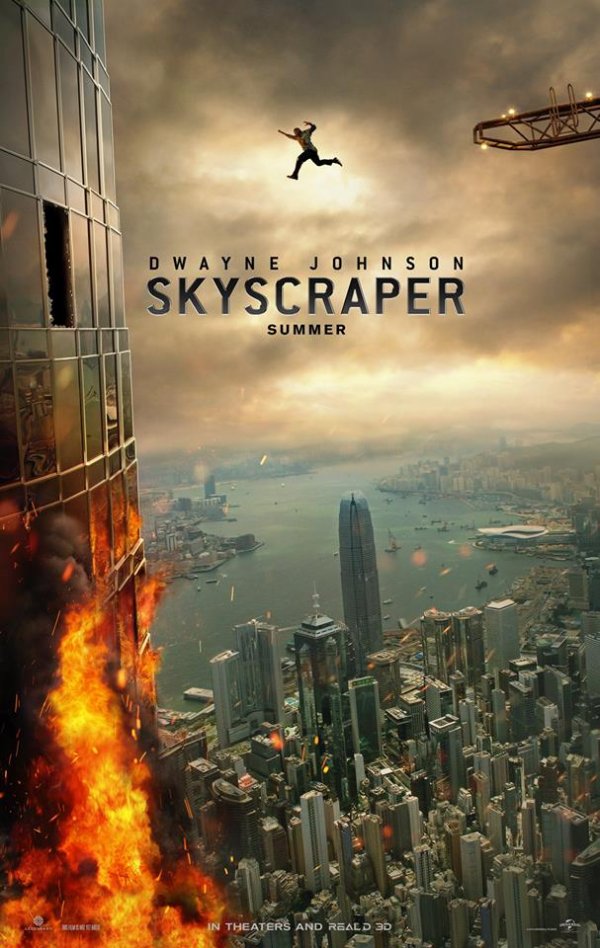}
\end{center}
\caption{\footnotesize Universal Pictures' teaser poster for \textsf{Skyscraper}.}
\label{fig:poster}
\end{wrapfigure}
\textsf{Skyscraper} is an action film starring Dwayne Johnson scheduled for release on July 13, 2018. In order to elevate the interest in the film, Universal Pictures released in February a teaser poster (seen in Figure \ref{fig:poster}) and a trailer. Both focus around an entertaining feat: an insane jump of the protagonist from an ultra-tall crane towards a broken window of a skyscraper. Not much is revealed in the released short trailer, but it got everyone  talking: Is the jump possible? 

I decided that the scene offers a great teaching moment from which students can benefit. It can be worked as  a Fermi problem.
And with the film not yet released, a little discussion about the feat can only make us demand more. I hence offer this article as a class activity to (a) understand the use of Fermi problems in deciding how to validate or refute the realism in a particular  scene and (b) provide an interesting way to expand on the standard projectile motion.

A general introduction on Fermi problems (also known as back-of-the-envelope problems) and a collection of them can be found in several sources: \cite{Swartz,vonBaeyer,WA,Weinstein}.
In relation to movies, I have published already a related article in this journal with the collaboration of the late professor R. A. Llewellyn
\cite{EL}. And additional articles analyzing scenes from Hollywood movies have appeared in many other articles and talks. Some are \cite{EL2,EL3,EL4}. Finally, an assessment of the method can be seen in \cite{ELMW}.

\section{Theoretical Investigation}

\subsection*{The set-up of the problem}

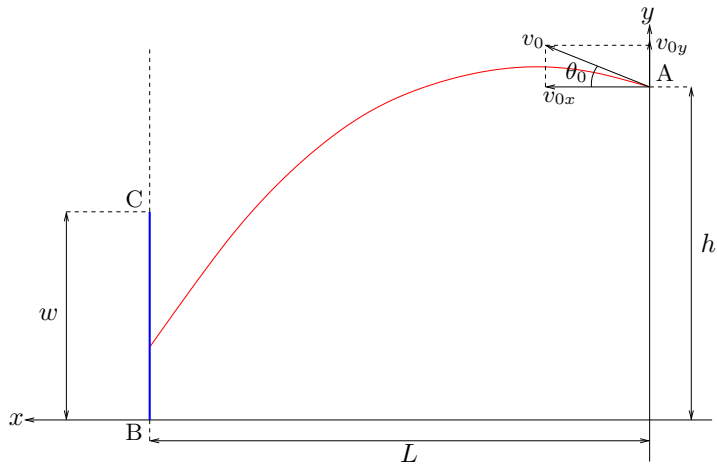
\begin{figure}[ht!]
\begin{center}
\setlength{\unitlength}{1mm}
\begin{picture}(100,50)
     \put(0,-5){\includegraphics[width=10cm]{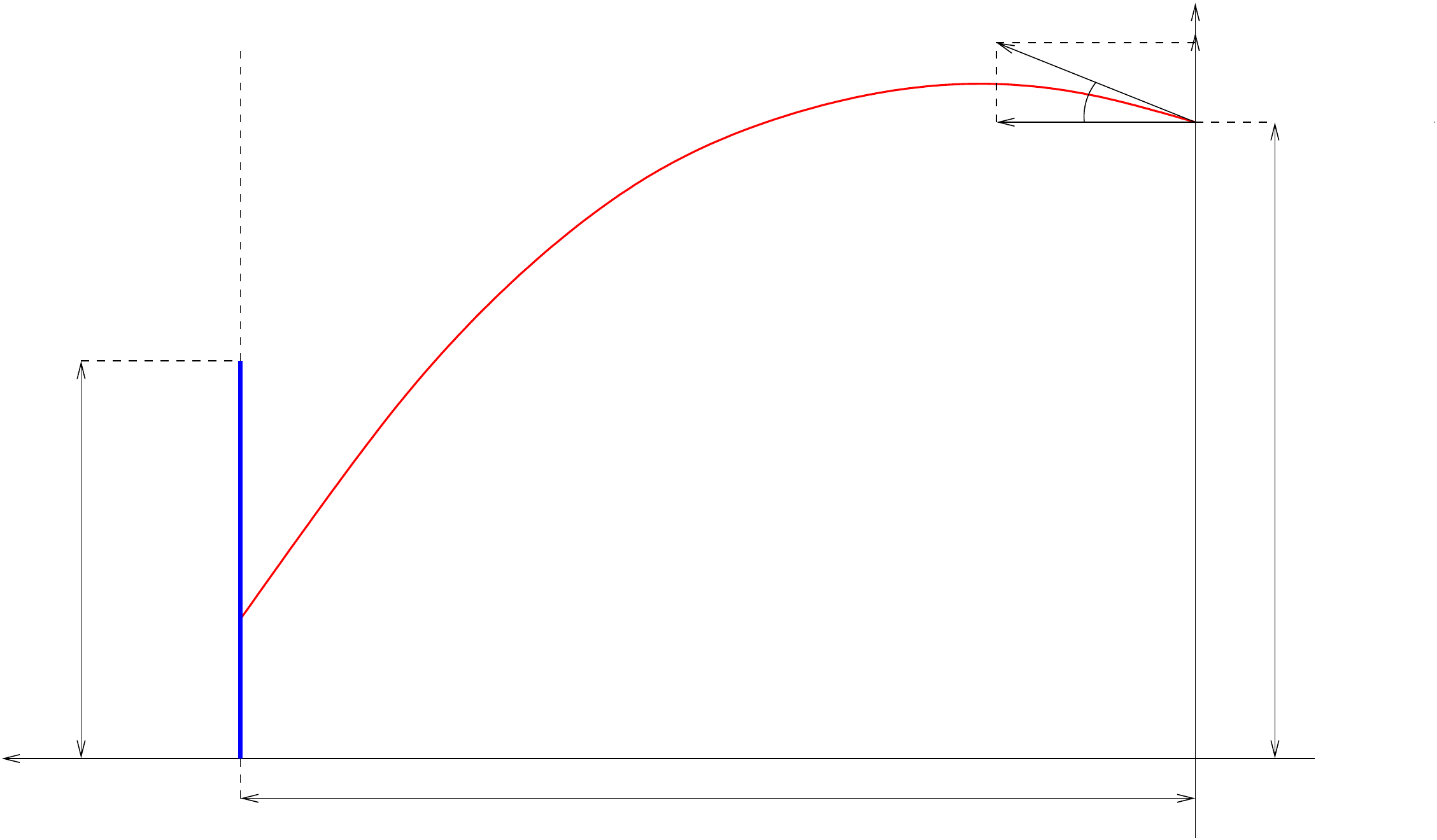}}
     \put(50,-5){\small $L$}
     \put(82,54){\small $y$}
     \put(-2,0){\small $x$}
     \put(2,14){\small $w$}
     \put(90,23){\small $h$}
     \put(69,43){\footnotesize $v_{0x}$}
     \put(84,50){\footnotesize $v_{0y}$}
     \put(66.2,50.5){\footnotesize $v_0$}
     \put(72,46){\footnotesize $\theta_0$}
     \put(84,45.7){\footnotesize A}
     \put(13.5,-2){\footnotesize B}
     \put(13.5,29){\footnotesize C}
\end{picture}
\end{center}
\caption{\footnotesize The set-up of the scene looks as a standard textbook problem: A projectile is fired from height $h$ above the
              ground. For what values of the speed will reach point A or B? The only difference from the standard problems is that a 
              student must extract the given quantities from the visual images using scaling arguments.}
\label{fig:SetUp}
\end{figure}

Figure \ref{fig:SetUp} shows the set-up of the problem as it would appear in a standard physics textbook: A projectile is fired from point
A  at height $h$ with initial speed $v_0$ at an angle $\theta_0$ above the horizontal line. For what values of $v_0$ and $\theta_0$ (equivalently, $v_{0x}$ and $v_{0y}$) will the projectile hit a vertical wall located at $x=L$ between the  ground point B  and the point C at height $w$? In this form, point $A$ represents the edge of the crane and BC the window of the skyscraper.

The difference from a standard problem is that the student will have to decide what is known by relying on the visual information provided by the director in the film.  In particular, one can compare the size $\ell$ of the protagonist with the
size of the window $w$, the distance $L$ between the crane and the skyscraper, and the vertical distance $h$ from the bottom of the
window to the crane. In doing so, since the action is shown through a viewing angle that favors visual pleasure
but it is not optimal  for the collection of data, we will  approximate the various quantities  in favor of the director to avoid any criticism for unfairness. Since the legs of the protagonist are not fully extended, we will assume that $\ell\simeq$1.5 meters. The reader can also easily check with the use of a ruler
that $L\sim10\ell$ (it appears to be more), $h\sim5\ell$ (it appears to be less than half of $L$), $w\sim2\ell$ (it appears to be less; a 3-meters window appears to be a decent assumption). In fact, since the protagonist is an extended object and not a point, we may assume that point C is the highest possible point  which he should reach in order to be able to pass through the window.

\subsection*{Analytical Calculations}

Given the initial speeds $v_{0x}$, $v_{0y}$ and using the coordinate as shown in Figure \ref{fig:SetUp}, 
the equations describing the projectile motion are the well known ones:
\begin{align}
     x     &= v_{0x} \, t ,                                       \label{eq:PM1}\\
     y     &=  h + v_{0y}\,t - {1\over2} \, g \, t^2,   \label{eq:PM2}\\
     v_y &= v_{0y} -g\, t                                       \label{eq:PM3}.
\end{align}
Eliminating the time from the first two equations, we find the equation $y=f(x)$ describing the trajectory:
$$
     y    =  h +  {v_{0y} \over v_{0x}} \, x - {g\over 2v^2_{0x}} \,  x^2 .
$$
Of course, it is an arc of a parabola.
At the location of the window,  the coordinate $x$ has the value $x=L$. The range of allowed speeds to perform the feat well is found by the demand that the protagonist lands inside the window. That is, when he reaches the skyscraper, his height $y$ from the ground is between the $y$-values of the top point C ($y=w$) and the bottom point A ($y=0$).  Hence,
\begin{equation}
    w   \ge   h +  {v_{0y} \over v_{0x}} \, L - {g\over 2v^2_{0x}} \,  L^2  \ge 0 .
\label{eq:BasicIneq}
\end{equation}
These two inequalities define the range of possible values of the speeds.  

\begin{figure}[ht!]
\begin{center}
\setlength{\unitlength}{1mm}
\begin{picture}(80,60)
    \put(0,-5){\includegraphics[width=8cm]{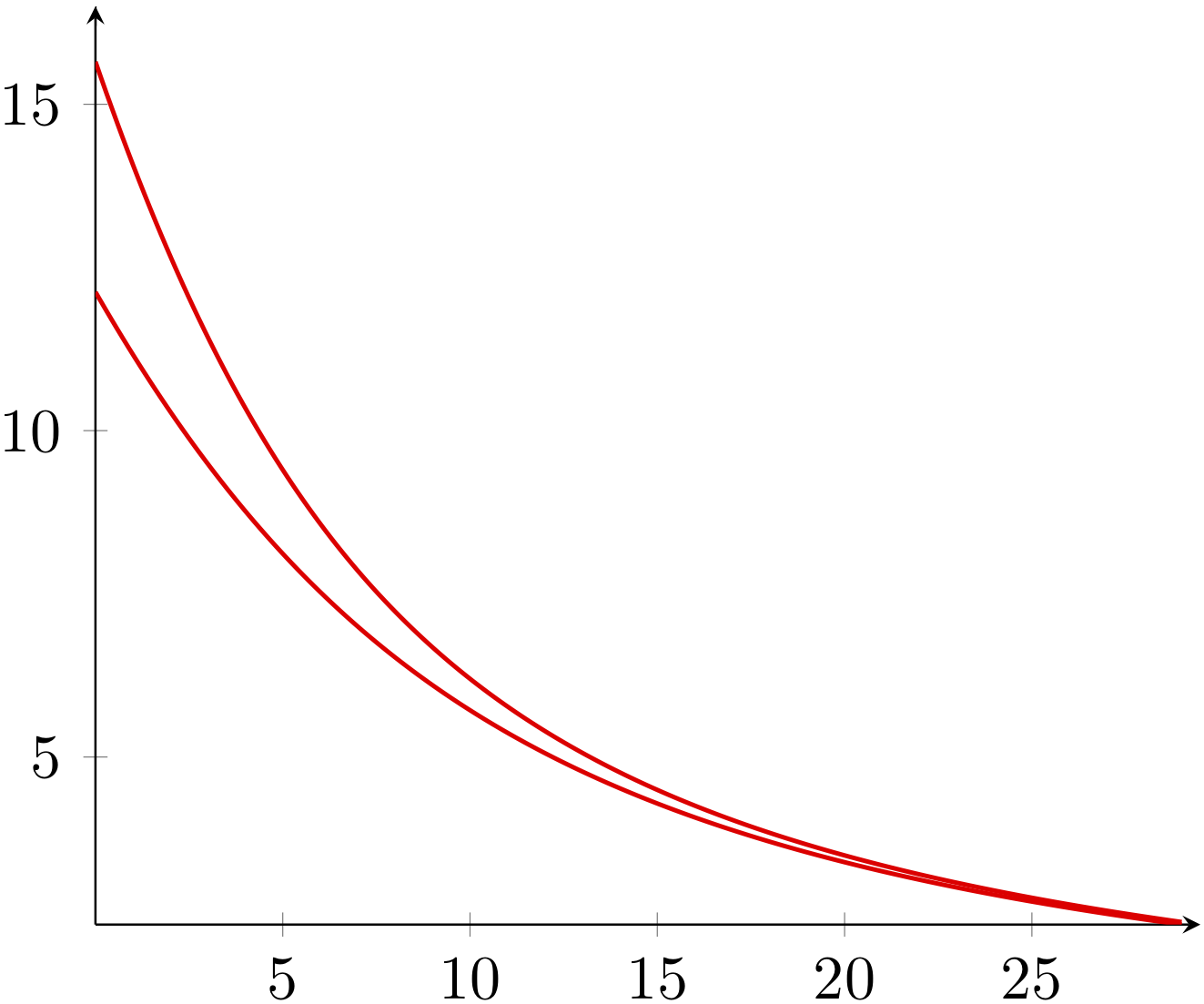}}
   \put(0,61){$v_{0x}$}
     \put(78,-3){$v_{0y}$}
\end{picture}
\end{center}
\caption{\footnotesize The required range for the initial speeds $v_{0x}$, $v_{0y}$ is the area between the two red curves. 
              The graph is drawn to scale.}
\label{fig:Range}
\end{figure}

In particular, for a given vertical speed, the horizontal speed  must fall in the interval
\begin{equation}
    {g L \over v_{0y}+\sqrt{v^2_{0y}+2gh}} \le v_{0x}  \le {g L\over v_{0y}+\sqrt{v^2_{0y}+2g(h-w)}} .
\label{eq:IneqX}
\end{equation}
From this, we can conclude:
\begin{itemize}
\item The smaller the height $w$ of the window, the smaller the allowed range for the horizontal speed. For a zero window height, $w=0$, the only option is
$$
       v_{0x} =  {g L \over v_{0y}+\sqrt{v^2_{0y}+2gh}} .
$$ 
\item The larger the distance $L$ to travel, the larger the speed $v_{0x}$ must be. (The lower bound increases.) In the limit
$L\to+\infty$, we also have $v_{0x}\to+\infty$.
\item The larger the vertical speed, the smaller the required horizontal speed. (The upper bound decreases.) In the limit
$v_{0y}\to+\infty$, we have $v_{0x}\to0$.
\end{itemize}
Using the length $\ell$ to rewrite equation \eqref{eq:IneqX}, we have
\begin{equation}
    {10g\ell\over v_{0y}+\sqrt{v^2_{0y}+10g\ell}} \le v_{0x}  \le {10g\ell\over v_{0y}+\sqrt{v^2_{0y}+6g\ell}} .
\label{eq:IneqX2}
\end{equation}
The allowed range of values is seen in Figure \ref{fig:Range}.

On the other hand, given  the horizontal speed in inequality  \eqref{eq:BasicIneq}, the vertical speed  must fall in the interval
\begin{equation}
    {g L^2-2hv^2_{0x} \over 2Lv_{0x}}    \le   v_{0y}   \le  {g L^2-2(h-w)v^2_{0x} \over 2Lv_{0x}} .
\label{eq:IneqY}
\end{equation}
From this, we can conclude:
\begin{itemize}
\item The smaller the height $w$ of the window, the smaller the allowed range for the vertical speed. For a zero window height, $w=0$, the only option is
$$
       v_{0y} =   {g L^2-2hv^2_{0x} \over 2Lv_{0x}} .
$$ 
\item The larger the distance $L$ to travel, the larger the speed $v_{0y}$ must be. (The lower bound increases.) In the limit
$L\to+\infty$, we also have $v_{0y}\to+\infty$.
\item The smaller the horizontal speed, the larger the required vertical speed. (The lower bound inreases.) In the limit
$v_{0x}\to0$, we have $v_{0y}\to+\infty$.
\end{itemize}
Using the length $\ell$ to rewrite inequality \eqref{eq:IneqY}, we have
\begin{equation}
   {10g\ell- v^2_{0x} \over 2v_{0x}}    \le   v_{0y}   \le  {50g\ell-3v^2_{0x} \over 10v_{0x}} .
\label{eq:IneqY2}
\end{equation}
This range is still that shown in Figure  \ref{fig:Range}.

\subsection*{Is the Feat Possible?}

For no initial vertical speed, $v_{0y}=0$, inequality \eqref{eq:IneqX2} requires that 
$$
     \sqrt{10g\ell} \le v_{0x}  \le {10\over \sqrt{6}} \, \sqrt{g\ell} 
     \Rightarrow
     12.124\, \text{m/s} \le v_{0x}  \le 15.652\, \text{m/s}.
$$
A professional sprinter runs 100 meters in (less than) 10 seconds. This is an average speed of  about 10 m/s. However, the average speed is not  the athlete's speed at the end of the race. Assuming a uniform acceleration, that would be 20 m/s. Hence, the above range is not completely out of question. However, the sprinter has a distance of 100 meters to achieve his final speed at an acceleration of 2 m/s$^2$. In the trailer, the protagonist probably has about 20 meters. (Later in the trailer, it shows him running again; the length may be 
a little more.) Assuming that he can perform as well as a professional sprinter, he will only be able to reach a speed close to 9 m/s. Therefore,  the feat is not humanly possible without vertical speed. 

The protagonist of the film has lost one of his legs and uses a prosthetic one. You may  question if a person with a prosthetic leg can perform as well as a fully functioning person.  Oscar Pistorius \cite{OP}, who has two prosthetic legs, is the world-record holder of the 100-meter  sprint with a time of 10.91  seconds.  This works out to an average speed of
9.166 m/s. Although about 10\% slower than that of a non-handicapped athlete, it is still a great performance. The final speed achieved at the end of the race  is 18.333 m/s. So, having a prosthetic leg is not a problem at all. Running  along the crane which  is made of steel pieces pinned together to form a lattice structure with big gaps in between is a way more serious problem. Fortunately, the director has
taken care of this problem; later in the trailer, he shows  us that there is a metallic carpet runner on the crane which our hero uses as a runaway.

Having decided that it is impossible to do the jump with zero initial vertical speed, let's now explore the possibility to perform the feat with a non-vanishing vertical speed. In fact, from the trailer we  clearly see that this is what the director has in  his mind.
Again from the trailer, we can see that the peak point of the trajectory is achieved approximately 
about a second after the jump. Then, from equation \eqref{eq:PM3}, we conclude  that  $v_{0y}$ is of the order  
10  m/s. Have we then reached   a final conclusion that the feat is not possible? Not yet. Achieving this  speed along the horizontal direction is solely an issue of human performance; it can only be achieved by running. Achieving it along the vertical direction however is not a matter of running; it is achieved by an impulse from the crane along the vertical direction. Athletes utilize impulses from the ground to thrust themselves high in almost all sports. 
At the edge of the crane, the protagonist  pushes with his foot down on the crane for a brief interval $\Delta t$. The crane reacts: it applies the vertical (exactly opposite) force  $F$  during the same time interval, thus producing the impulse
$J=F \, \Delta t$ which, in turn, produces the (change of) momentum $m v_{0y}$:
\begin{equation}
      F \, \Delta t = m \, v_{0y} .
\label{eq:Impulse}
\end{equation}
According to Wikipedia \cite{DJ}, Dwayne Johnson's mass is 118 kg. Hence, we can compute the force $F$ on him from this equation, 
if we 
know how long the action-reaction between him and the crane lasted. Data from athletes show that the faster the movement of the sport, the shorter the contact time $\Delta t$ with the ground.  Such an interaction always lasts a  fraction of a second;
for sprinters this is about 0.1 s. For high and long jumpers this time increases slightly since the athlete has to press down on the ground
longer  to get a good impulse. This duration does depend on technique too. 
Overall, accepting the value $\Delta t$=0.1 s for our hero is a reasonable assumption. Inserting the numbers in the equation above, we conclude that the force $F$ on him from the crane is 11,800 Newtons. 
We can approximate the protagonist's foot as a rectangle with size  a total surface area $A$ of 200 cm$^2$ for the foot. 
We  may further assume that, to get the impulse, he pushes down with the entire area of his  foot to get a vertical boost. Then, the pressure applied by the crane on the foot is
$$
     P = {F\over A} = 0.59 \, \text{MPa}.
$$
The average compressive pressure which the bones can withstand is many MPa (the exact number is not important here); the previous exerted pressure is well below this limit. The weakest place of his leg is the ankle joint. It is known \cite{Burdett} though that the ankle can withstand 13 times a person's weight while running. Writing equation \eqref{eq:Impulse} in the form
$$
     F= mg \, {v_{0y}\over g\,\Delta t},
$$
we see that the applied force on the foot is $\lambda= v_{0y} / (g\Delta t)$ times the protagonist's weight. Inserting the assumed values of the quantities, we find the numerical value of $\lambda$ to be close to 10. Hence, within reasonable limits; so far, we have found no serious objections.

Substituting the values of the various quantities in the right-hand side of equation \eqref{eq:PM2}, we see that the peak point
of the trajectory   has $y$=12.6 meters.  This is 2.6 meters above the crane. The world record for high jump  is
2.45 meters. And athletes use the Fosbury flop to achieve it since, although the bar stands at a height, the center-of-mass reaches a
lower height. In the current feat, as portrayed in the film, our protagonist goes airborne in a vertical stance. That is, he truly needs to raise
his center-of-mass by 2.6 meters.
However, watching the trailer, we can  see  that the protagonist  does not actually go very high above the crane (follow his center-of-mass located approximately at his waist); he raises only by a fraction of $\ell$=1.5 meters. Although, it is still impressive for a vertical jump while keeping the body straight up, we have discovered a `mathematicovisual' inconsistency: the visual motion does not match the mathematical model it corresponds to. My assumption is that  the director has created the scene such that the actual jump appears realistic --- the jump is similar to those seen during basketball games. It could be a stuntman's jump superimposed on the crane-building background.

It already appears a doomed jump  for $v_{0y}$=10 m/s, but let's keep going to better understand all aspects.  
Inequality \eqref{eq:IneqX2} requires
\begin{align*}
           5.716  \, \text{m/s} \le v_{0x}  \le     6.198 \, \text{m/s} .
\end{align*}
Actually, this range is not bad at all.  For a well trained person, it is achievable ---  even if one of his legs is prosthetic and even if the
available length to run is no more than 20 meters.     Equation  \eqref{eq:PM1} now requires that, within a second, the projectile  travels a distance  between 5.716 and 6.198 meters. But in the trailer it is shown  to be about $\ell$=1.5 meters. This is certainly 
another big visual versus mathematical disagreement --- one by a factor of 4. We become more disappointed once we realize that, in his descent from the peak point, the hero will cross the horizontal plane at the same height with  the crane at a distance $R$ satisfying
\begin{align*}
  11.432 \, \text{m} \le R \le 12.396 \, \text{m} .
\end{align*}
The current world record for long jump is 8.95 meters. This indicates that achieving the combination of  values for $v_{0x}$ and $v_{0y}$  as discussed above  is not trivial at all. The required values break the world records for high and long jump by a big margin.
Even if we assume that our hero had been training to break the world records in the two sports exactly by the above margins, he would 
not be able to do so \textit{simultaneously}. Our discussion
assumes that the two initial speeds can be achieved independently of each other. However,  the construction of the human body enforces constraints which, in turn, make the two speeds tightly related. If the athlete focuses on optimizing the horizontal speed, it is not easy to optimize the vertical speed too. Vice versa, if the athlete focuses on optimizing the vertical speed, it is not easy to optimize the horizontal speed too. Depending on the situation, the athlete decides which component of the two is to be optimized. 
Consequently, our conclusion is that achieving the values of $v_{0x}$ and $v_{0y}$ as discussed above is beyond human capabilities.

We reached a dead end after we decided to interpret the time of the scene as real time. However, let's now adopt the attitude that the time shown on screen is not real time. In doing so, we can resolve the visual-mathematical inconsistencies we discovered previously if we assume that the peak is reached at an earlier time --- hence, the motion is shown at a slower rate.  Allowing our hero to accelerate as fast as a professional sprinter, he decides to utilize the entire length of the crane available to him to reach  maximum horizontal speed $v_{0x}$=9 m/s.  Inequality \eqref{eq:IneqY2} then  requires a vertical speed $v_{0y}$ in the range  3.667 m/s and 5.467 m/s. In this case, the peak point is achieved within 0.37 s and 0.56 s, implying a peak elevation 0.686 m and 1.525 m above the crane. This is actually quite realistic. Running at 9 m/s and jumping about 70 cm   resembles the performance of professional basketball players. Achieving both of the required speeds simultaneously is consistent with human capabilities.  
We must therefore conclude that, within the various uncertainties in the approximations, the feat shown in the released trailer
and poster of the \textsf{Skyscraper} is actually possible!

\section{Epilogue}

In recent times, Hollywood has opted for extreme entertainment at the expense of logic and science. Usually, most of the feats seen
in the films are impossible or just plain wrong when analyzed with simple science. Movies like \textsf{Contact} and 
\textsf{Interstellar} which are based on correct science are a scarce commodity in Hollywood despite their box office success.
Yet, from time to time, we uncover scenes in movies which, although they portray an extreme feat driven only by the intend to entertain 
the hard core action fans, are possible. The scene just released for \textsf{Skyscraper} proves to be along these lines.
Made for breathtaking fun,  no sane person would actually attempt to perform it no matter how elevated his adrenaline is.
A simple back-of-the-envelope analysis shows that the feat can be performed for a variety of values for $v_{0x}$, $v_{0y}$.  Although for  most of these values the feat is impossible for humans, there is choice which is consistent with a trained human.
Therefore,  our hero can perform the feat successfully and save his family. 
When the film is finally released and more precise visual data can be harnessed, an updated calculation may be done to check if the current conclusion stands. Hopefully, the result is not accidental and the
director has used a science advisor to work out some of the  details and make the film's stunts realistic.

\section*{Addendum}

\begin{figure}[h!]
\begin{center}
\includegraphics[height=7cm]{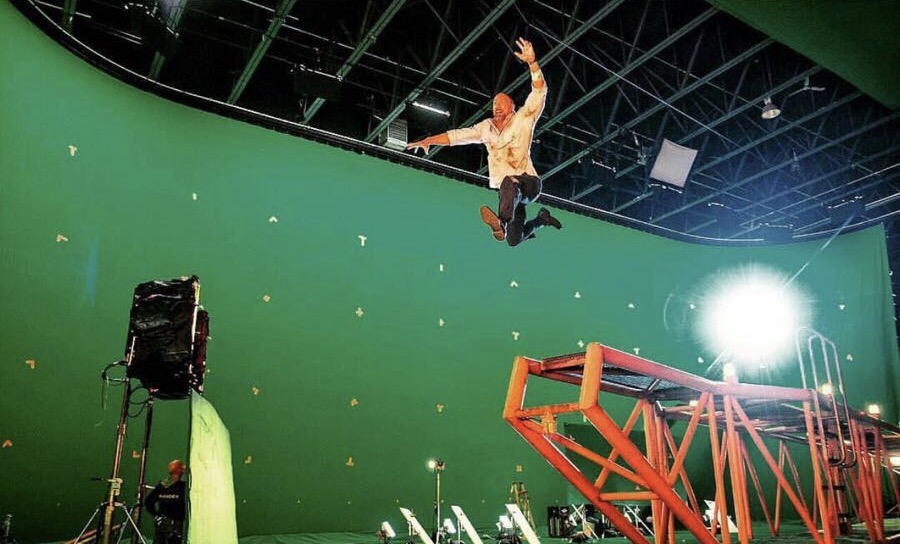}
\end{center}
\caption{\footnotesize In this photo found in IMDb, its is clearly seen that the jump shown in the movie is an actual jump done
              by Dwayne Johnson in front of a green screen. The city background was added later. As much as we can say from this 
              photo, there appears to be no cable attached to Johnson to enhance and/or modify his jump.}
\label{fig:jump}
\end{figure}

 After I had written the present article and an abridged version \cite{Efthimiou} was approved for publication in \textit{Physics Education}, I discovered on IMDb the photo shown in Figure \ref{fig:jump}.  From this we are ensured that the jump shown in the film  is a real one; it was
performed by  Dwayne Johnson himself. 
To his credit, although Johnson is 46 years old (born May 2, 1972 according to Wikipedia \cite{DJ}), his  abilities  match those of the much younger basketball players. We also indirectly conclude that the jump is shown at a slower rate.

I hope that educators, especially high school physics teachers, will find in this article a perfect example to demonstrate to students how physics can be used in forensic science: If Dwayne Johnson had presented the clip as proof of his supernatural abilities, we would have argued that there was nothing supernatural about it. Simple back-of-the-envelope calculations revealed immediately that the jump was impossible for humans as it requires abilities exceeding Olympic-Games-level  human performance.  The calculations  also pointed out that the jump could be based on an actual jump and  clever digital editing could have been used to exaggerate it, making it look as supernatural.    Of course, at the end, no editing can deceive science.  The clip --- this clip or any other similar clip --- contains information which, in the hands of a scientist, can be used as a powerful tool towards the discovery of the truth. \textit{And this is exactly what physicists do: They look at `clips' in the world that surrounds them and from the data they collect, they infer the real workings of the universe.}

\section*{Acknowledgements}

I would like to thank Jamal Khayat for providing comments and feedback on the article.


\end{document}